\documentclass[11pt,a4paper]{article}

\usepackage[utf8]{inputenc}
\usepackage[T1]{fontenc}
\usepackage{times}
\usepackage{amsmath,amssymb,amsfonts}
\usepackage{amsthm}
\usepackage{graphicx}
\usepackage{booktabs}
\usepackage{multirow}
\usepackage{xcolor}
\usepackage{tikz}
\usetikzlibrary{shapes,arrows,positioning,fit,calc,backgrounds}
\usepackage{listings}
\usepackage{algorithm}
\usepackage{algpseudocode}
\usepackage{hyperref}
\usepackage{cleveref}
\usepackage[margin=1in]{geometry}
\usepackage{enumitem}
\usepackage{caption}
\usepackage{subcaption}
\usepackage{longtable}
\usepackage{array}

\newtheorem{theorem}{Theorem}[section]
\newtheorem{definition}[theorem]{Definition}

\newtheorem{proposition}[theorem]{Proposition}

\lstdefinestyle{yaml}{
    basicstyle=\ttfamily\footnotesize,
    breaklines=true,
    frame=single,
    numbers=left,
    numberstyle=\tiny\color{gray},
    keywordstyle=\color{blue},
    stringstyle=\color{red},
    commentstyle=\color{green!60!black},
    showstringspaces=false,
    columns=fullflexible,
}

\lstdefinestyle{prompt}{
    basicstyle=\ttfamily\footnotesize,
    breaklines=true,
    frame=single,
    backgroundcolor=\color{gray!10},
    showstringspaces=false,
    columns=fullflexible,
}

\title{\textbf{4D-ARE: Bridging the Attribution Gap in LLM Agent Requirements Engineering}}

\author{
    Bo Yu\textsuperscript{1} \quad
    Lei Zhao\textsuperscript{1} \\[0.5em]
    \textsuperscript{1}Tencent \\[0.3em]
    \texttt{\{evenyu, sanshizhao\}@tencent.com}
}

\date{January 2026 \\ \small{Preprint. Under review.}}

\begin{document}

\maketitle

\begin{abstract}
We deployed an LLM agent with ReAct reasoning and full data access. It executed flawlessly, yet when asked ``Why is completion rate 80\%?'', it returned metrics instead of causal explanation. The agent knew \emph{how} to reason but we had not specified \emph{what} to reason about.

This reflects a gap: runtime reasoning frameworks (ReAct, Chain-of-Thought) have transformed LLM agents, but design-time specification---determining what domain knowledge agents need---remains under-explored.

We propose \textbf{4D-ARE} (4-Dimensional Attribution-Driven Agent Requirements Engineering), a preliminary methodology for specifying attribution-driven agents. The core insight: decision-makers seek attribution, not answers. Attribution concerns organize into four dimensions (Results $\to$ Process $\to$ Support $\to$ Long-term), motivated by Pearl's causal hierarchy. The framework operationalizes through five layers producing artifacts that compile directly to system prompts.

We demonstrate the methodology through an industrial pilot deployment in financial services. 4D-ARE addresses \emph{what} agents should reason about, complementing runtime frameworks that address \emph{how}. We hypothesize systematic specification amplifies the power of these foundational advances. This paper presents a methodological proposal with preliminary industrial validation; rigorous empirical evaluation is planned for future work.
\end{abstract}

\noindent\textbf{Keywords:} Agent Requirements Engineering, LLM Agents, Causal Attribution, Goal-Oriented Requirements Engineering, Prompt Engineering, Decision Support Systems

\section{Introduction}
\label{sec:introduction}

\subsection{A Failure That Worked Perfectly}

Consider a deployment scenario: an LLM agent designed to help bank managers review relationship manager performance. The agent has access to all relevant data---deposit completion rates, customer visit logs, product penetration metrics, staffing reports. It uses ReAct-style reasoning to think step by step. By every technical measure, it works beautifully---the reasoning-action loop executes flawlessly.

Yet the agent failed to deliver value.

When a regional manager asked ``Why is Eastern region's completion rate only 80\%?'', the agent responded: ``Eastern region's deposit completion rate is 80\%. Visit frequency is 4.2 per week, which is lower than other regions. Product penetration is 24\%. The team should work to improve these metrics.''

Technically correct (accurate data retrieval, valid reasoning steps). Practically insufficient. The manager didn't need more numbers---she needed to understand \emph{why} performance was lagging and \emph{what} to do about it. The agent answered the literal question while completely missing the point.

This failure crystallized a problem we had been circling for months: \textbf{the agent knew how to retrieve and reason, but we had never specified what it should reason about}. The fault was ours, not the framework's.

\subsection{The Attribution Gap}

The failure above illustrates a fundamental gap in how intelligent systems support decision-makers:

\begin{quote}
\textbf{Decision-makers ask about outcomes, but what they truly seek is attribution.}
\end{quote}

When a manager asks ``Why is completion rate 80\%?'', they are not requesting a data lookup. They want causal explanation: which process failed? what resource was missing? what strategic factor is at play? They want an \emph{attribution chain} that connects observed outcomes to actionable causes.

An agent that traces ``completion rate is 80\% $\leftarrow$ low product penetration in high-value segment $\leftarrow$ insufficient visit coverage $\leftarrow$ elevated competitive pressure from 3 new market entrants'' has provided attribution. It has given the manager something to act on.

Current agent design methodologies do not produce such responses. They focus on \emph{how} agents reason at runtime (ReAct, Chain-of-Thought, planning), while design-time specification---determining \emph{what} agents should reason about---has received comparatively less attention. Runtime reasoning frameworks assume a well-specified agent. In practice, that assumption is almost never met.

\subsection{The Complementary Challenge: Design-Time Specification}

The research community has invested heavily in runtime reasoning. ReAct~\cite{yao2023react} teaches agents to interleave thinking and acting. Chain-of-Thought~\cite{wei2022cot} elicits step-by-step inference. Tree-of-Thoughts~\cite{yao2023tot} explores branching solution spaces. Tool-augmented approaches~\cite{schick2023toolformer,shen2023hugginggpt} extend agent capabilities. These are real advances.

But they address a different problem from the one we tackle here.

Runtime frameworks answer: \emph{How should the agent think during execution?}

They assume someone has already answered: \emph{What should the agent think about?}

That assumption is the gap. In practice, agent designers face an unaddressed challenge: translating domain expertise into agent specification. What questions should the agent monitor? What causal relationships should structure its reasoning? What boundaries should constrain its behavior? These design-time decisions determine whether an agent produces attribution or noise.

The result is what we call the \textbf{promptware crisis}: agent development degrades to trial-and-error prompt engineering, implicit knowledge that lives in developers' heads, and deployed agents that fail in predictable ways---incomplete explanations, boundary violations, missed causal connections. The root cause is not inadequate runtime reasoning. It is absent design-time specification.

\subsection{4D-ARE: Attribution-Driven Agent Specification}

This paper presents \textbf{4D-ARE} (4-Dimensional Attribution-Driven Agent Requirements Engineering), a methodology for specifying LLM agents grounded in causal attribution logic.

The core insight: decision-maker concerns organize into four dimensions that form a causal chain.

\begin{figure}[htbp]
\centering
\begin{tikzpicture}[
    node distance=0.8cm,
    dimbox/.style={rectangle, draw=black, fill=blue!10, minimum width=2.8cm, minimum height=1.2cm, align=center, font=\small},
    arrow/.style={->, thick, >=stealth}
]
\node[dimbox] (results) {\textbf{RESULTS}\\``What outcomes?''};
\node[dimbox, right=of results] (process) {\textbf{PROCESS}\\``What actions?''};
\node[dimbox, right=of process] (support) {\textbf{SUPPORT}\\``What resources?''};
\node[dimbox, right=of support] (longterm) {\textbf{LONG-TERM}\\``What context?''};

\draw[arrow] (results) -- (process);
\draw[arrow] (process) -- (support);
\draw[arrow] (support) -- (longterm);

\node[below=0.3cm of results, font=\scriptsize, align=center] {Observable\\metrics};
\node[below=0.3cm of process, font=\scriptsize, align=center] {Controllable\\activities};
\node[below=0.3cm of support, font=\scriptsize, align=center] {Configurable\\capabilities};
\node[below=0.3cm of longterm, font=\scriptsize, align=center] {Environmental\\factors};
\end{tikzpicture}
\caption{The Four Dimensions. When Results show a gap, attribution traces backward through Process, Support, and Long-term to identify causes.}
\label{fig:four-dimensions}
\end{figure}
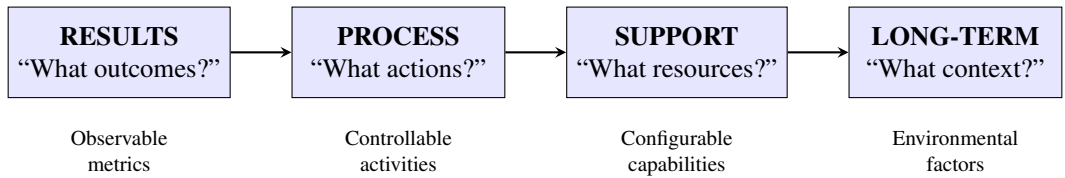

This structure is not arbitrary. We motivate it from Pearl's causal hierarchy and control theory (Section~\ref{sec:why-four}), arguing that attribution models satisfying causal completeness, actionability separation, and temporal coherence naturally organize into these four dimensions. The framework is theoretically grounded, though we acknowledge the dimensional structure is a useful conceptual organization rather than a mathematical necessity.

4D-ARE operationalizes this insight through five layers:

\begin{table}[htbp]
\centering
\caption{The Five-Layer Architecture}
\label{tab:five-layers}
\begin{tabular}{@{}lll@{}}
\toprule
\textbf{Layer} & \textbf{Specifies} & \textbf{Produces} \\
\midrule
1. Question Inventory & What the agent perceives & Perception scope \\
2. Attribution Model & How the agent traces causality & Reasoning structure \\
3. Data Mapping & What data the agent accesses & Tool configuration \\
4. Dual-Track Logic & When to interpret vs.\ recommend & Output policies \\
5. Boundary Constraints & What the agent must not do & Safety guardrails \\
\bottomrule
\end{tabular}
\end{table}

Each layer produces explicit artifacts that compile directly to system prompts. This closes the specification gap: domain expertise translates systematically to deployed agent.

\subsection{What We Contribute}

We evaluate 4D-ARE through an industrial pilot: designing a Performance Attribution Agent for relationship manager review at a commercial bank. This pilot demonstrates:

\begin{itemize}[leftmargin=*]
    \item \textbf{Systematic translation}: How domain expertise maps to agent specification through the five layers
    \item \textbf{Concrete artifacts}: YAML specifications that compile directly to system prompts
    \item \textbf{Boundary enforcement}: How explicit constraints prevent inappropriate agent behavior
    \item \textbf{Practical effectiveness}: Expert usability, attribution completeness, and boundary compliance
\end{itemize}

The implication: runtime reasoning frameworks assume a well-specified agent. 4D-ARE provides that specification. They are complementary, and we argue that specification is the higher-leverage, currently under-addressed investment.

\subsection{Contributions and Epistemic Status}

\textbf{Note on claims:} The four-dimensional structure is motivated by theory but not derived from it (see Proposition~\ref{thm:necessity}). Observations from deployment are preliminary and should be treated as hypotheses for future investigation, not validated findings.

\textbf{Conceptual:}
\begin{enumerate}[leftmargin=*]
    \item Identification of the attribution gap as a practical challenge in decision-support agent design
    \item A four-dimensional framework motivated by causal reasoning principles (Proposition~\ref{thm:necessity} presents this as a hypothesis, not a theorem)
    \item Extension of Goal-Oriented Requirements Engineering concepts for LLM agents
\end{enumerate}

\textbf{Practical:}
\begin{enumerate}[leftmargin=*,resume]
    \item Five-layer specification architecture with concrete YAML artifacts
    \item Structured elicitation methods for tacit attribution logic
\end{enumerate}

\textbf{Experiential:}
\begin{enumerate}[leftmargin=*,resume]
    \item Industrial pilot demonstrating end-to-end methodology
    \item Preliminary observations (not rigorous evaluation) on usability and effectiveness
\end{enumerate}

\subsection{Scope and Organization}

4D-ARE addresses LLM agents for decision support in enterprise contexts---agents that support human decision-makers, operate in domains with causal structure, require explainable outputs, and must respect operational boundaries. We position this work at the intersection of agent-oriented software engineering and requirements engineering, contributing a systematic methodology where previously only ad-hoc practices existed.

Section~\ref{sec:related} positions our work against existing approaches. Section~\ref{sec:framework} motivates the four-dimensional structure. Section~\ref{sec:operationalization} details the five-layer methodology. Section~\ref{sec:case-study} presents the industrial pilot. Section~\ref{sec:discussion} discusses implications. Complete specification artifacts are provided in the Appendix for full reproducibility.

\section{Related Work}
\label{sec:related}

LLM agent design draws from multiple traditions. We examine each and identify what is missing.

\subsection{Runtime Reasoning: Foundational Advances}

The research community has achieved transformative progress on \textbf{runtime reasoning}---how agents think during execution. These advances form the foundation upon which practical agent systems are built.

\textbf{Chain-of-Thought and Extensions.} Wei et al.'s Chain-of-Thought prompting~\cite{wei2022cot} was a landmark discovery: LLMs solve complex problems dramatically better when generating intermediate reasoning steps. This insight spawned a rich family of techniques. Self-Consistency~\cite{wang2023selfconsistency} improves robustness by sampling multiple reasoning paths. Tree-of-Thoughts~\cite{yao2023tot} enables systematic exploration of branching solutions.

\textbf{ReAct and Tool Use.} Yao et al.'s ReAct~\cite{yao2023react} elegantly unified reasoning and acting, showing that interleaving thought and action produces more capable and interpretable agents. This framework has become a cornerstone of agent design. Toolformer~\cite{schick2023toolformer} teaches models when to call external APIs. HuggingGPT~\cite{shen2023hugginggpt} orchestrates multiple models as tools.

\textbf{Planning.} LLM-based planning~\cite{huang2022lmplanning,wang2023plansolve} enables sophisticated task decomposition. AutoGPT~\cite{richards2023autogpt} explores end-to-end autonomous completion.

These frameworks have made LLM agents genuinely useful. The agent we described in Section 1 leveraged ReAct's reasoning-action loop and faithfully executed its logic. The problem was not with ReAct---it performed exactly as designed. The problem was \emph{upstream}: we had not specified what domain knowledge the agent needed. ReAct excels at executing reasoning; domain expertise must be provided separately.

\textbf{The opportunity:} Runtime reasoning frameworks are powerful engines that need fuel. 4D-ARE provides a methodology for systematically specifying the domain knowledge these frameworks reason over.

\subsection{Requirements Engineering: Right Idea, Wrong Era}

Goal-Oriented Requirements Engineering (GORE)~\cite{dardenne1993kaos,yu1997istar,bresciani2004tropos} provides systematic methods for specifying software systems. KAOS models goals and obstacles. i* models actor dependencies. Tropos guides agent-oriented development.

These frameworks capture important insights: goals decompose hierarchically; agents have capabilities and limitations; obstacles must be anticipated. But they were designed for deterministic software agents. LLM agents are different:

\begin{itemize}[leftmargin=*]
    \item GORE assumes agents behave consistently. LLM agents are probabilistic.
    \item GORE models obstacles as exceptions. LLM agents need ``must not do'' specifications as primary constraints.
    \item GORE produces code specifications. LLM agents need prompt specifications.
\end{itemize}

Agent-Oriented Software Engineering~\cite{jennings2000aose,wooldridge2009intro,padgham2004prometheus,zambonelli2003gaia} extends these ideas for multi-agent systems. But AOSE assumes agents are \emph{implemented} as code. LLM agents are \emph{configured} as prompts. The artifacts don't transfer.

\textbf{The gap:} RE methodology exists, but not for LLM agents. We need to adapt, not adopt.

\subsection{Knowledge Engineering: The Tacit Problem}

CommonKADS~\cite{schreiber2000commonkads} provides comprehensive methods for knowledge-based system development. Ontology engineering~\cite{guarino1998ontology,noy2001ontology101} formalizes domain concepts.

The problem: these approaches assume knowledge can be made \emph{fully explicit}. For decision support agents, the critical knowledge is \emph{attribution logic}---how experts connect outcomes to causes. This logic is often tacit. When asked ``How do you know low completion rate is caused by inventory problems?'', experts say ``I just see the pattern.'' They cannot articulate the causal chain.

CommonKADS has no method for eliciting tacit causal reasoning. Ontologies capture what exists, not how to reason about causality.

\textbf{The gap:} KE assumes explicit knowledge. Attribution logic is partly tacit. New elicitation methods are needed.

\subsection{Prompt Engineering: Technique Without Methodology}

Prompt engineering has produced many techniques~\cite{schulhoff2024promptreport}. Few-shot examples. Role prompting. Output formatting. DSPy~\cite{khattab2023dspy} separates signatures from implementations and enables automatic prompt optimization.

These approaches address \emph{how} to write effective prompts. They do not address \emph{what} to include for a given domain. A practitioner facing a new domain has techniques for prompt construction but no methodology for determining prompt content.

\textbf{The gap:} Prompt technique is well-developed. Prompt content methodology is absent.

\subsection{Specification-Driven Development: Process Without Content}

Recent work has explored specification-driven approaches for AI-assisted development. GitHub's spec-kit\footnote{\url{https://github.com/github/spec-kit}} provides workflows for ``spec-driven development''---defining requirements before implementation. Similar tools structure the development \emph{process} (specify $\to$ plan $\to$ implement).

These approaches address \emph{how to organize development}. They do not address \emph{what domain knowledge} an agent needs. spec-kit helps developers articulate what they want to build; 4D-ARE helps practitioners articulate what an agent should know about a domain.

The distinction is:
\begin{itemize}[leftmargin=*]
    \item \textbf{Process specification} (spec-kit, DSPy): How to structure development/optimization
    \item \textbf{Content specification} (4D-ARE): What domain knowledge enables useful agent behavior
\end{itemize}

A practitioner could use spec-kit to organize their development process \emph{and} 4D-ARE to specify the agent's domain knowledge. The approaches are complementary at different levels of the design stack.

\textbf{The gap:} Development process tools exist. Domain knowledge specification methodology is absent.

\subsection{Causal Attribution in AI Systems}

Recent work has explored causal attribution in multi-agent and LLM contexts, but primarily for \emph{runtime explanation} rather than \emph{design-time specification}.

\textbf{Runtime Attribution.} MACIE~\cite{macie2025} combines structural causal models with Shapley values to explain collective agent behavior after execution. A2P Scaffolding~\cite{a2p2025} applies structured counterfactual reasoning (Abduct-Act-Predict) for automated failure attribution in multi-agent systems. These approaches answer ``why did the agent do X?'' but do not address ``what should the agent reason about?''

\textbf{Collaborative Sensemaking.} Collaborative Causal Sensemaking~\cite{ccs2025} proposes a research agenda for human-AI teams to jointly construct and revise shared causal models. This is closer to 4D-ARE's goals but focuses on training regimes and interaction policies rather than systematic specification.

\textbf{Requirements Engineering for LLM Systems.} MARE~\cite{jin2024mare} proposes multi-agent collaboration across the entire requirements engineering process (elicitation, modeling, verification, specification). Elicitron~\cite{elicitron2024} uses LLM agents to simulate diverse users for design requirements elicitation. ROPE~\cite{rope2024} emphasizes requirement-oriented prompt engineering to help humans articulate clearer requirements for LLMs. These address \emph{how to gather requirements}, not \emph{how to structure agent knowledge around causal attribution}.

\textbf{The gap:} Existing causal attribution work focuses on runtime explanation or training. Design-time specification of attribution logic remains unaddressed.

\subsection{Summary: Six Opportunities, One Methodology}

\begin{table}[htbp]
\centering
\caption{How 4D-ARE Complements Existing Traditions}
\label{tab:gaps}
\begin{tabular}{@{}p{3.0cm}p{3.0cm}p{4.5cm}@{}}
\toprule
\textbf{Tradition} & \textbf{Opportunity} & \textbf{4D-ARE Contribution} \\
\midrule
Runtime Reasoning & Needs domain specification & Provides specification methodology \\
Requirements Engineering & Designed for code agents & Adapts for LLM agents \\
Knowledge Engineering & Assumes explicit knowledge & Elicits tacit attribution \\
Prompt Engineering & Technique without content & Content methodology \\
Spec-Driven Development & Process without domain knowledge & Domain knowledge specification \\
Causal Attribution & Runtime not design-time & Design-time attribution specification \\
\bottomrule
\end{tabular}
\end{table}

4D-ARE complements these traditions through a single insight: \emph{structure agent specification around causal attribution}. The four dimensions provide the knowledge structure. The five layers provide the methodology. The output is a deployable system prompt that empowers runtime frameworks like ReAct and CoT to deliver domain-appropriate reasoning.

\section{The 4D-ARE Framework}
\label{sec:framework}

\subsection{Core Insight: The Attribution Gap}

We begin with an observation that, while simple, has profound implications for agent design:

\begin{quote}
\textbf{Decision-makers ask about outcomes, but what they truly seek is attribution.}
\end{quote}

Traditional business intelligence systems answer ``what''---sales completion rate is 80\%. But decision-makers actually need to know ``why''---which process failed---and ``how''---what action to take next. This gap between \emph{outcome reporting} and \emph{causal attribution} represents a fundamental challenge that current agent design methodologies fail to address.

\begin{definition}[Attribution Gap]
\label{def:attribution-gap}
Let $Q$ be a decision-maker's query and $R$ be an agent's response. The \emph{attribution gap} $\mathcal{G}(Q, R)$ is the semantic distance between:
\begin{itemize}
    \item The \emph{surface-level answer} $R_s$ (e.g., ``completion rate is 80\%'')
    \item The \emph{attribution-complete answer} $R_a$ (e.g., ``completion rate is 80\% because inventory backlog in Region A caused by delayed new product launch'')
\end{itemize}
\end{definition}

Current prompting methodologies (ReAct, CoT, function calling) optimize agent \emph{reasoning at runtime} but provide no systematic method for specifying \emph{what the agent should reason about}. The 4D-ARE framework addresses this design-time gap.

\subsection{Motivation: Why Four Dimensions?}
\label{sec:why-four}

A natural question arises: why four dimensions? Is this choice principled or arbitrary? We motivate the four-dimensional structure from two independent perspectives---Pearl's causal hierarchy and control-theoretic decomposition---and show they suggest a similar four-layer organization. We emphasize that this is motivation, not derivation: we do not claim four dimensions are mathematically necessary, only that they represent a useful design choice grounded in established theory.

\subsubsection{Motivation from Pearl's Causal Hierarchy}

Pearl's \emph{Ladder of Causation}~\cite{pearl2018book} distinguishes three levels of causal reasoning:

\begin{enumerate}
    \item \textbf{Association} (Seeing): $P(Y|X)$ --- What is the outcome given observation?
    \item \textbf{Intervention} (Doing): $P(Y|do(X))$ --- What happens if I act?
    \item \textbf{Counterfactual} (Imagining): $P(Y_x|X', Y')$ --- What would have happened if...?
\end{enumerate}

When a decision-maker asks ``Why is completion rate 80\%?'', they are not asking for association (Level 1). They seek \emph{intervention targets}: which actions would change the outcome? This is Level 2 reasoning. But intervention requires understanding \emph{causal mechanisms}---which leads to Level 3 counterfactual reasoning about what \emph{could have been different}.

We observe that organizational causation operates through a specific mechanism:

\begin{proposition}[Organizational Causal Chain]
\label{prop:causal-chain}
In organizational decision-making, outcomes $Y$ are generated through a four-stage causal process:
\begin{equation}
\text{Environment} \to \text{Capabilities} \to \text{Actions} \to \text{Outcomes}
\end{equation}
where each stage causally influences the next, and earlier stages have slower rates of change.
\end{proposition}

This proposition is grounded in organizational theory. Outcomes (quarterly sales) result from actions (customer visits, cross-selling). Actions are constrained by capabilities (staffing, tools, skills). Capabilities develop within environmental conditions (market dynamics, competitive pressure, regulatory context).

Mapping to Pearl's hierarchy:
\begin{itemize}
    \item \textbf{Outcomes} = What we observe (Association level)
    \item \textbf{Actions} = What we can intervene on (Intervention level)
    \item \textbf{Capabilities} = What enables/constrains intervention (Counterfactual: ``if we had more staff...'')
    \item \textbf{Environment} = What we cannot control but must account for (Counterfactual: ``if competitors hadn't entered...'')
\end{itemize}

The counterfactual level naturally splits into two: factors within organizational control (Capabilities/Support) and factors outside organizational control (Environment/Long-term). This split is not arbitrary---it reflects the fundamental distinction between \emph{actionable} and \emph{contextual} causes that decision-makers must navigate.

\subsubsection{Motivation from Control Theory}

An independent derivation comes from control theory. Consider the standard state-space model:

\begin{equation}
\begin{aligned}
x_{t+1} &= f(x_t, u_t, w_t) \quad \text{(state dynamics)} \\
y_t &= g(x_t, v_t) \quad \text{(observation)}
\end{aligned}
\end{equation}

where $y_t$ is observed output, $x_t$ is internal state, $u_t$ is control input, $w_t$ is process noise (environmental disturbance), and $v_t$ is observation noise.

For a decision-maker observing $y_t$ and asking ``why?'', the causal decomposition is:

\begin{equation}
y_t \leftarrow x_t \leftarrow u_{t-1} \leftarrow \pi(x_{t-1}) \leftarrow w_{t-1}
\end{equation}

where $\pi$ is the policy (how actions are chosen given state). This gives us four distinct causal contributors:

\begin{table}[htbp]
\centering
\caption{Control-Theoretic Derivation of Four Dimensions}
\label{tab:control-derivation}
\begin{tabular}{@{}lll@{}}
\toprule
\textbf{Control Theory} & \textbf{4D-ARE} & \textbf{Decision-Maker Question} \\
\midrule
Output $y_t$ & Results ($\mathcal{D}_R$) & ``What happened?'' \\
Control input $u_t$ & Process ($\mathcal{D}_P$) & ``What did we do?'' \\
State $x_t$ & Support ($\mathcal{D}_S$) & ``What did we have to work with?'' \\
Disturbance $w_t$ & Long-term ($\mathcal{D}_L$) & ``What external factors affected us?'' \\
\bottomrule
\end{tabular}
\end{table}

\subsubsection{Convergence and Interpretation}

The two perspectives---from causal inference and control theory---suggest a similar four-layer structure. This convergence is suggestive but not conclusive: it indicates that four dimensions capture something useful about how outcomes are generated in organizational systems, but does not prove that four is the uniquely correct number.

\begin{proposition}[Dimensional Organization---Hypothesis]
\label{thm:necessity}
We hypothesize that an attribution model for organizational decision-making benefits from satisfying:
\begin{enumerate}
    \item \textbf{Causal completeness}: Every outcome gap can be traced to contributing factors
    \item \textbf{Actionability separation}: Factors are distinguished by decision-maker's ability to intervene
    \item \textbf{Temporal coherence}: Factors are ordered by rate of change (fast to slow)
\end{enumerate}
A model satisfying these properties naturally organizes into four categories corresponding to: observable outcomes, controllable actions, configurable resources, and environmental context.
\end{proposition}

\textbf{Argument.} Condition (1) requires distinguishing outcomes from their causes---minimally two layers. Condition (2) requires separating causes into those directly controllable (actions) vs.\ indirectly controllable (resources)---splitting causes into two layers. Condition (3) further separates resources (changeable within planning horizon) from environment (exogenous within planning horizon). Thus: outcomes + actions + resources + environment = four categories. 

This argument motivates four dimensions but does not prove uniqueness. Alternative decompositions with different granularity remain possible.

\textbf{A note on epistemic status.} The argument above is conceptual rather than mathematically rigorous. The three conditions are motivated by practical decision-making needs, not derived from axioms. Alternative categorizations with different granularity (three or five dimensions) are conceivable. We argue that four dimensions represent a useful balance---coarse enough to be memorable, fine enough to be actionable---but do not claim mathematical uniqueness.\footnote{Why not three? Collapsing Support and Long-term loses the actionability distinction---decision-makers need to know whether a causal factor is something they can change (hire more staff) or something they must adapt to (competitor entry). Why not five? One might propose splitting Process into ``planning'' and ``execution,'' but this violates temporal coherence---both operate at the same timescale. Additional granularity can exist \emph{within} dimensions without requiring new dimensions.}

\subsubsection{Alignment with Established Frameworks}

The four-dimensional structure aligns with established management frameworks, providing external validation:

\begin{table}[htbp]
\centering
\caption{Alignment with Established Frameworks}
\label{tab:framework-alignment}
\begin{tabular}{@{}lcccc@{}}
\toprule
\textbf{Framework} & \textbf{$\mathcal{D}_R$} & \textbf{$\mathcal{D}_P$} & \textbf{$\mathcal{D}_S$} & \textbf{$\mathcal{D}_L$} \\
\midrule
Balanced Scorecard~\cite{kaplan1992bsc} & Financial & Internal Process & Learning \& Growth & Customer/Market \\
PESTEL Analysis & --- & --- & Internal & External \\
McKinsey 7S & Goals & Systems/Style & Staff/Skills/Structure & Strategy \\
\bottomrule
\end{tabular}
\end{table}

The Balanced Scorecard is particularly instructive: Kaplan and Norton~\cite{kaplan1992bsc} derived their four perspectives from extensive empirical study of what metrics executives actually use. Their four perspectives map closely to our four dimensions, suggesting that the 4D structure captures something fundamental about how decision-makers think.

\subsection{Formal Definition of the Four Dimensions}

Having established theoretical necessity, we now formally define each dimension.

\begin{definition}[4D Attribution Model]
\label{def:4d-model}
An attribution model $\mathcal{A}$ is a four-tuple:
\begin{equation}
\mathcal{A} = \langle \mathcal{D}_R, \mathcal{D}_P, \mathcal{D}_S, \mathcal{D}_L \rangle
\end{equation}
where:
\begin{itemize}
    \item $\mathcal{D}_R$ (Results): Observable outcome metrics---the ``what'' that triggers attribution queries
    \item $\mathcal{D}_P$ (Process): Controllable actions---directly manipulable by decision-makers
    \item $\mathcal{D}_S$ (Support): Configurable resources---changeable within planning horizon but not immediately
    \item $\mathcal{D}_L$ (Long-term): Environmental context---exogenous factors that constrain but cannot be controlled
\end{itemize}
\end{definition}

\begin{definition}[Attribution Dependency]
\label{def:attribution-dependency}
The four dimensions form a directed acyclic graph with causal direction:
\begin{equation}
\mathcal{D}_L \xrightarrow{\text{constrains}} \mathcal{D}_S \xrightarrow{\text{enables}} \mathcal{D}_P \xrightarrow{\text{produces}} \mathcal{D}_R
\end{equation}
Attribution traces this chain \emph{backward}: when $\mathcal{D}_R$ shows a gap, the agent asks ``which $\mathcal{D}_P$ failed?'' then ``which $\mathcal{D}_S$ was missing?'' then ``which $\mathcal{D}_L$ changed?''
\end{definition}

\textbf{Principle (Attribution Completeness).}
\label{thm:attribution-completeness}
An agent response $R$ is \emph{attribution-complete} with respect to query $Q$ if and only if $R$ addresses all dimensions in the backward causal path from the queried dimension.

\textbf{Rationale.} By Proposition~\ref{thm:necessity}, any outcome gap has contributing factors in each of the four dimensions. An explanation that omits any dimension leaves the decision-maker without actionable insight for that causal pathway. Specifically:
\begin{itemize}
    \item Omitting $\mathcal{D}_P$: Decision-maker knows \emph{what} happened but not \emph{what to change}
    \item Omitting $\mathcal{D}_S$: Decision-maker knows \emph{what to change} but not \emph{what enables the change}
    \item Omitting $\mathcal{D}_L$: Decision-maker may attempt changes that are futile given environmental constraints
\end{itemize}
Thus, completeness requires all four dimensions.

\subsection{Dimension Specifications and Agent Authority}

Each dimension has distinct characteristics regarding agent behavior:

\begin{table}[htbp]
\centering
\caption{Dimension Characteristics and Agent Behavior}
\label{tab:dimension-behavior}
\begin{tabular}{@{}lcccc@{}}
\toprule
\textbf{Characteristic} & \textbf{$\mathcal{D}_R$} & \textbf{$\mathcal{D}_P$} & \textbf{$\mathcal{D}_S$} & \textbf{$\mathcal{D}_L$} \\
\midrule
Rate of change & Fast (daily) & Medium (weekly) & Slow (monthly) & Very slow (quarterly+) \\
Controllability & Observable & Direct control & Indirect control & No control \\
Agent interpretation & None & Rule-based & None & None \\
Agent recommendation & None & Specific & Open-ended & None \\
\bottomrule
\end{tabular}
\end{table}

The key insight is that \textbf{agent authority decreases as we move from Process to Long-term}. The agent can confidently interpret and recommend for Process (actionable, fast-changing, data-rich). For Support, recommendations must be open-ended (requires human judgment on resource allocation). For Long-term, the agent provides context only (strategic decisions are beyond agent scope).

This graduated authority is not arbitrary---it follows from the actionability separation principle in Proposition~\ref{thm:necessity}.

\subsection{The Attribution Chain as Reasoning Scaffold}

\begin{figure}[htbp]
\centering
\begin{tikzpicture}[
    node distance=0.5cm,
    dimbox/.style={rectangle, draw=black, fill=blue!5, minimum width=11cm, minimum height=1.4cm, align=left, font=\small},
    arrow/.style={->, thick, >=stealth, color=blue!60},
    whylabel/.style={font=\scriptsize\itshape, color=blue!60}
]
\node[dimbox] (dr) {
    \textbf{$\mathcal{D}_R$ (Results)}: ``Annual completion rate is 80\%''
};
\node[dimbox, below=of dr] (dp) {
    \textbf{$\mathcal{D}_P$ (Process)}: ``Inventory backlog severe; cross-sell conversion below threshold''
};
\node[dimbox, below=of dp] (ds) {
    \textbf{$\mathcal{D}_S$ (Support)}: ``Staffing ratio inadequate; campaign coverage at 67\%''
};
\node[dimbox, below=of ds] (dl) {
    \textbf{$\mathcal{D}_L$ (Long-term)}: ``3 new competitors entered; digital adoption accelerating''
};

\draw[arrow] (dr.south) -- (dp.north) node[midway, right, whylabel] {WHY? (What action failed?)};
\draw[arrow] (dp.south) -- (ds.north) node[midway, right, whylabel] {WHY? (What resource was missing?)};
\draw[arrow] (ds.south) -- (dl.north) node[midway, right, whylabel] {WHY? (What context changed?)};

\end{tikzpicture}
\caption{The Attribution Chain. Each ``WHY?'' traces backward through the causal hierarchy, from observable outcomes to environmental context.}
\label{fig:attribution-chain}
\end{figure}
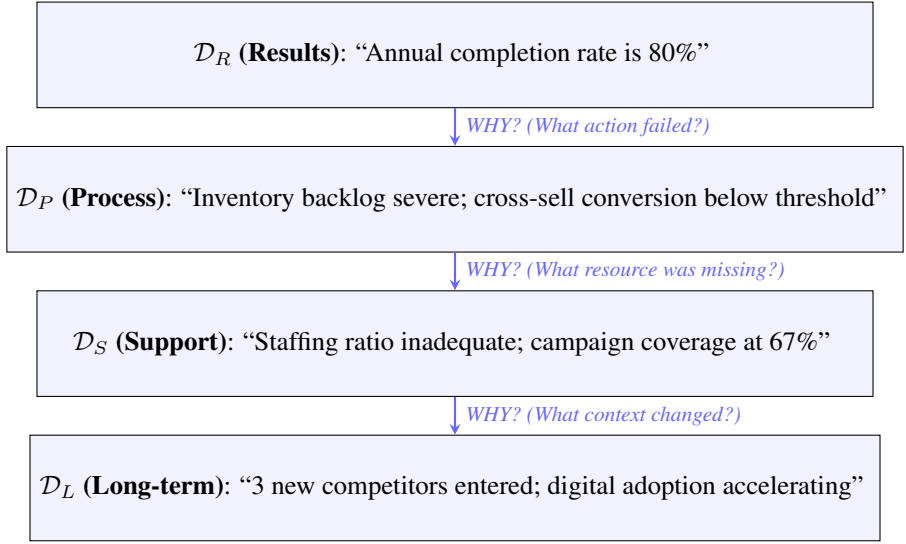

\begin{definition}[Attribution Trace]
\label{def:attribution-trace}
Given a query $Q$ targeting dimension $\mathcal{D}_i$, the \emph{attribution trace} $\mathcal{T}(Q)$ is the sequence of dimension transitions:
\begin{equation}
\mathcal{T}(Q) = \mathcal{D}_i \to \mathcal{D}_{i+1} \to \cdots \to \mathcal{D}_L
\end{equation}
The agent's system prompt encodes this trace as a mandatory reasoning protocol.
\end{definition}

\subsection{Extending GORE for LLM Agents}

We position 4D-ARE as a principled extension of Goal-Oriented Requirements Engineering (GORE)~\cite{dardenne1993kaos,yu1997istar}:

\begin{table}[htbp]
\centering
\caption{From GORE to 4D-ARE}
\label{tab:gore-extension}
\begin{tabular}{@{}p{2.8cm}p{3.2cm}p{5cm}@{}}
\toprule
\textbf{GORE Concept} & \textbf{4D-ARE Extension} & \textbf{Rationale} \\
\midrule
Goal hierarchy & Attribution chain & Goals decompose by \emph{causal dependency}, not just refinement \\
Softgoals & Interpretation rules & Quality attributes become explicit reasoning constraints \\
Obstacles & Boundary constraints & Failure modes become ``must not do'' specifications \\
Deterministic agents & LLM Agent + Tools & Probabilistic behavior requires explicit guardrails \\
\bottomrule
\end{tabular}
\end{table}

The key extension is recognizing that for LLM agents supporting decision-makers, the goal hierarchy must encode \emph{how to attribute outcomes to causes}---because the agent's value lies in surfacing causal relationships, not just achieving goals.

\section{Operationalizing 4D-ARE}
\label{sec:operationalization}

The theoretical foundation established in Section~\ref{sec:framework}---the four dimensions and their causal dependencies---provides the conceptual framework for attribution-driven design. This section addresses operationalization: a five-layer architecture that translates theoretical constructs into concrete specification artifacts.

\subsection{The Five-Layer Architecture}

4D-ARE operationalizes through a five-layer architecture that translates business domain knowledge into agent specifications:

\begin{figure}[htbp]
\centering
\begin{tikzpicture}[
    layer/.style={rectangle, draw=black, minimum width=12cm, minimum height=1.2cm, align=center, font=\small},
    arrow/.style={<->, thick, >=stealth, color=gray}
]
\node[layer, fill=green!10] (l1) at (0,0) {\textbf{Layer 1: QUESTION INVENTORY} --- ``What the agent perceives''};
\node[layer, fill=green!15] (l2) at (0,1.4) {\textbf{Layer 2: ATTRIBUTION MODEL} --- ``How the agent reasons about causality''};
\node[layer, fill=green!20] (l3) at (0,2.8) {\textbf{Layer 3: DATA MAPPING} --- ``What data sources the agent accesses''};
\node[layer, fill=green!25] (l4) at (0,4.2) {\textbf{Layer 4: DUAL-TRACK LOGIC} --- ``How the agent interprets and recommends''};
\node[layer, fill=green!30] (l5) at (0,5.6) {\textbf{Layer 5: BOUNDARY CONSTRAINTS} --- ``What the agent must NOT do''};

\end{tikzpicture}
\caption{The Five-Layer Specification Architecture. Each layer produces artifacts that compile to system prompt components.}
\label{fig:five-layers}
\end{figure}
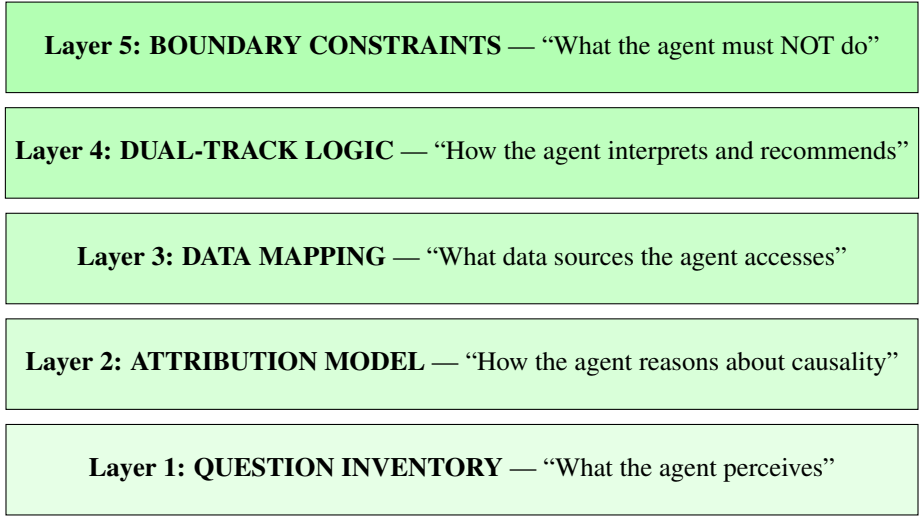

Each layer corresponds to a specific component of the agent's system prompt. While presented sequentially, practitioners may iterate between layers as specifications mature---for example, data mapping challenges (Layer 3) may reveal gaps in the question inventory (Layer 1).

\subsection{Layer 1: Question Inventory (Agent Perception Scope)}

\textbf{Objective:} Exhaustively enumerate decision-maker concerns to define the agent's ``eyes''---what it can perceive.

\textbf{Method:} Structured elicitation with domain experts using the 4D framework as an interview scaffold:

\begin{table}[htbp]
\centering
\caption{Elicitation Prompts by Dimension}
\label{tab:elicitation}
\begin{tabular}{@{}p{2cm}p{5cm}p{4cm}@{}}
\toprule
\textbf{Dimension} & \textbf{Elicitation Prompt} & \textbf{Example Questions} \\
\midrule
Results & ``What outcomes do you track to judge success?'' & Completion rates, revenue targets \\
Process & ``What operational indicators signal problems?'' & Inventory levels, conversion rates \\
Support & ``What resources must be in place for success?'' & Market coverage, team capacity \\
Long-term & ``What strategic factors affect future performance?'' & Product pipeline, competitive landscape \\
\bottomrule
\end{tabular}
\end{table}

\textbf{Output:} Question set $\mathcal{Q} = \{q_1, q_2, \ldots, q_n\}$ classified by dimension.

\textbf{Specification artifact:}
\begin{lstlisting}[style=yaml]
question_inventory:
  results:
    - id: Q1
      question: "Monthly/annual completion rate by sales unit?"
      data_source: erp_sales_table
      
  process:
    - id: Q3
      question: "Cross-region scan rate status?"
      data_source: knowledge_base_file
      interpretation: required
      recommendation: required
\end{lstlisting}

\subsection{Layer 2: Attribution Model (Agent Reasoning Framework)}

\textbf{Objective:} Construct the causal dependency structure that guides agent reasoning.

\textbf{Method:} For each question in $\mathcal{Q}$, identify:
\begin{enumerate}
    \item Which dimension it belongs to
    \item What downstream questions it explains (for $\mathcal{D}_R, \mathcal{D}_P, \mathcal{D}_S$)
    \item What upstream questions explain it (for $\mathcal{D}_P, \mathcal{D}_S, \mathcal{D}_L$)
\end{enumerate}

\textbf{Output:} Attribution graph $G = (V, E)$ where:
\begin{itemize}
    \item $V = \mathcal{Q}$ (questions as nodes)
    \item $E = \{(q_i, q_j) \mid q_j$ causally depends on $q_i\}$
\end{itemize}

\textbf{Example attribution chain:}
\begin{verbatim}
Q1 (Completion rate 80%)
  |---> Q4 (Inventory backlog)
         |---> Q8 (Key market development delayed)
                |---> Q10 (New product launch behind schedule)
\end{verbatim}

\textbf{Specification artifact:}
\begin{lstlisting}[style=yaml]
attribution_model:
  chains:
    - trigger: Q1  # Results gap detected
      trace: [Q4, Q8, Q10]  # Process -> Support -> Long-term
      
  dimension_mapping:
    results: [Q1, Q2]
    process: [Q3, Q4, Q5, Q6, Q7]
    support: [Q8, Q9]
    longterm: [Q10, Q11]
\end{lstlisting}

\subsection{Layer 3: Data Mapping (Agent Tool Specification)}

\textbf{Objective:} Define how the agent retrieves data for each question.

\textbf{Method:} For each $q \in \mathcal{Q}$, specify:
\begin{enumerate}
    \item Data source type (structured DB, knowledge base file, API)
    \item Query logic (SQL template, file naming convention, API endpoint)
    \item Update frequency and freshness requirements
\end{enumerate}

\textbf{Output:} Tool configuration for each question.

\textbf{Specification artifact:}
\begin{lstlisting}[style=yaml]
data_mapping:
  Q1:  # Sales completion
    source_type: database
    connection: erp_middleware
    query_template: |
      SELECT unit, month, actual/target as completion_rate
      FROM sales_summary
      WHERE year = {current_year}
    update_frequency: daily
    
  Q3:  # Cross-region scan rate
    source_type: knowledge_base
    file_pattern: "scan_rate_{YYYY}_{MM}.xlsx"
    update_frequency: monthly
    parse_method: excel_to_structured
\end{lstlisting}

\subsection{Layer 4: Dual-Track Logic (Agent Action Specification)}

\textbf{Objective:} Define both \emph{what the agent says} (interpretation) and \emph{what it suggests} (recommendation).

\textbf{Key insight:} Not all questions warrant interpretation or recommendation. The dual-track logic explicitly specifies agent behavior per question:

\begin{table}[htbp]
\centering
\caption{Dual-Track Logic by Dimension}
\label{tab:dual-track}
\begin{tabular}{@{}lccc@{}}
\toprule
\textbf{Question Type} & \textbf{Interpretation} & \textbf{Recommendation} \\
\midrule
Results ($\mathcal{D}_R$) & $\times$ None & $\times$ None \\
Process ($\mathcal{D}_P$) & $\checkmark$ Rule-based & $\checkmark$ Rule-based \\
Support ($\mathcal{D}_S$) & $\times$ None & $\sim$ Open-ended only \\
Long-term ($\mathcal{D}_L$) & $\times$ None & $\times$ None \\
\bottomrule
\end{tabular}
\end{table}

\textbf{Specification artifact:}
\begin{lstlisting}[style=yaml]
dual_track_logic:
  Q3:  # Cross-region scan rate
    interpretation:
      enabled: true
      rules:
        - condition: "rate > 0.15"
          output: "Elevated cross-region activity detected, market order risk"
        - condition: "rate <= 0.15"
          output: "Cross-region activity within normal range"
          
    recommendation:
      enabled: true
      rules:
        - condition: "rate > 0.15"
          output: "Consider anti-arbitrage measures to protect price integrity"
          
  Q4:  # Inventory status
    interpretation:
      enabled: true
      rules:
        - condition: "inflow > outflow"
          output: "Inventory accumulation detected"
        - condition: "inflow < outflow"
          output: "Inventory depletion detected"
          
    recommendation:
      enabled: true
      rules:
        - condition: "inflow > outflow"
          output: "Recommend moderating procurement pace"
        - condition: "inflow < outflow"
          output: "Recommend expedited replenishment to prevent stockout"
\end{lstlisting}

\subsection{Layer 5: Boundary Constraints (Agent Safety Specification)}

\textbf{Objective:} Define what the agent must NOT do---the ``red lines'' that prevent harmful or out-of-scope behavior.

\textbf{Insight:} For LLM agents, specifying boundaries is as important as specifying capabilities. Boundaries prevent:
\begin{enumerate}
    \item \textbf{Hallucination beyond data}: Agent fabricating causal relationships not supported by evidence
    \item \textbf{Scope creep}: Agent reasoning about domains outside its specification
    \item \textbf{Inappropriate confidence}: Agent making definitive claims where uncertainty exists
\end{enumerate}

\textbf{Specification artifact:}
\begin{lstlisting}[style=yaml]
boundary_constraints:
  global:
    - "Never fabricate data points not present in retrieved sources"
    - "Never provide recommendations for dimensions marked 'recommend: disabled'"
    - "Always caveat interpretations with data recency"
    
  per_question:
    Q1:
      - "Display data only; no interpretation of why targets were missed"
      - "No recommendations for improving completion rate"
      
    Q2:
      - "Out of scope for this agent deployment"
      
    Q8:
      - "Provide open-ended strategic suggestions only"
      - "Do not infer specific action plans"
\end{lstlisting}

\subsection{From Specification to System Prompt}

The five-layer specification translates directly into a structured system prompt:

\begin{lstlisting}[style=prompt]
# Agent System Prompt (Generated from 4D-ARE Specification)

## Role
You are a business intelligence agent for [Domain]. Your purpose is to help 
decision-makers understand performance through causal attribution.

## Perception Scope (Layer 1)
You monitor the following questions:
- [Q1]: Monthly/annual completion rate by sales unit
- [Q3]: Cross-region scan rate status
- [Q4]: Monthly inventory status by unit
...

## Reasoning Framework (Layer 2)
When answering questions, trace causality through dimensions:
- Results questions (Q1-Q2): Report facts, then trace to Process
- Process questions (Q3-Q7): Interpret data, recommend actions, trace to Support
- Support questions (Q8-Q9): Report status, provide open suggestions
- Long-term questions (Q10-Q11): Report information only

## Data Access (Layer 3)
For each question, use the following tools:
- Q1: query_erp_sales(unit, period)
- Q3: read_knowledge_base("scan_rate_{date}.xlsx")
...

## Interpretation & Recommendation Rules (Layer 4)
[Explicit rules per question as specified]

## Boundaries (Layer 5)
You must NOT:
- Interpret or recommend for Results dimension questions
- Provide detailed action plans for Support dimension questions
- Make claims about data you have not retrieved
...
\end{lstlisting}

\subsection{Methodology Summary: The 4D-ARE Process}

The complete methodology follows six steps:

\begin{enumerate}[leftmargin=*]
    \item \textbf{ELICIT}: Interview domain experts using 4D framework $\to$ Question Inventory (Layer 1)
    \item \textbf{MODEL}: Construct causal dependencies between questions $\to$ Attribution Model (Layer 2)
    \item \textbf{MAP}: Specify data sources and query logic per question $\to$ Data Mapping (Layer 3)
    \item \textbf{SPECIFY}: Define interpretation and recommendation rules $\to$ Dual-Track Logic (Layer 4)
    \item \textbf{CONSTRAIN}: Define boundaries and safety guardrails $\to$ Boundary Constraints (Layer 5)
    \item \textbf{GENERATE}: Compile five layers into System Prompt $\to$ Deployable Agent Specification
\end{enumerate}

\subsection{Key Contributions of This Section}

\begin{enumerate}[leftmargin=*]
    \item \textbf{Formalization}: We provide a formal definition of attribution-complete agent responses and the 4D attribution model that enables them.
    
    \item \textbf{Theoretical grounding}: We position 4D-ARE as a principled extension of GORE for LLM agents, addressing the unique challenges of probabilistic behavior and prompt-based configuration.
    
    \item \textbf{Operationalization}: The five-layer architecture provides a concrete, reproducible methodology for translating domain expert knowledge into agent specifications.
    
    \item \textbf{Specification artifacts}: Each layer produces explicit artifacts (YAML configurations) that directly translate to system prompt components---bridging the gap between requirements and implementation.
    
    \item \textbf{Boundary-first design}: Unlike approaches that focus primarily on what agents \emph{can} do, 4D-ARE emphasizes what agents \emph{must not} do---important for deploying agents in high-stakes business contexts.
\end{enumerate}

\section{Industrial Pilot: Deployment in Financial Services}
\label{sec:case-study}

We applied 4D-ARE to design a Performance Attribution Agent for relationship manager (RM) performance review at a commercial bank. This section reports on a \emph{pilot deployment} demonstrating the feasibility of the methodology in a real-world setting. To protect sensitive business intelligence, specific metrics and examples are constructed for demonstration purposes while reflecting realistic patterns. This section presents the deployment context, implementation process, and preliminary observations.

\subsection{Study Context}

\textbf{Representative scenario:} Quarterly performance reviews covering 200+ RMs across 15 branches. Current static reports answer ``what'' (metrics) but fail to explain ``why'' (attribution) or ``what next'' (recommendations).

\textbf{Agent objective:} Answer management's performance questions with full causal attribution, while respecting boundaries that reserve strategic decisions for human judgment.

\textbf{Deployment constraints:}
\begin{itemize}[leftmargin=*]
    \item Data sources: core banking system, CRM, manual branch reports
    \item Must not make personnel recommendations (HR policy)
    \item Must distinguish data-driven insights from strategic suggestions
\end{itemize}

\subsection{Method Implementation}

We present representative examples for each step. Complete specifications are provided in Appendix~\ref{appendix:specifications}.

\subsubsection{Step 1: Question Inventory Elicitation}

Structured interviews with COO, Branch Managers, and Data Analytics Team yielded 12 questions organized by dimension:

\begin{table}[htbp]
\centering
\caption{Question Inventory Summary}
\label{tab:questions-summary}
\begin{tabular}{@{}lcp{7cm}@{}}
\toprule
\textbf{Dimension} & \textbf{Count} & \textbf{Examples} \\
\midrule
Results ($\mathcal{D}_R$) & 2 & Q1: Deposit completion rate; Q2: AUM growth \\
Process ($\mathcal{D}_P$) & 5 & Q3: Visit frequency; Q4: Product penetration; Q5: Acquisition; Q6: Churn; Q7: Cross-sell \\
Support ($\mathcal{D}_S$) & 2 & Q8: Staffing adequacy; Q9: Campaign coverage \\
Long-term ($\mathcal{D}_L$) & 3 & Q10: Customer lifecycle value; Q11: Competitive pressure; Q12: Digital adoption \\
\bottomrule
\end{tabular}
\end{table}

\subsubsection{Step 2: Attribution Model Construction}

We mapped causal dependencies through facilitated workshops:

\begin{lstlisting}[style=yaml]
attribution_chains:
  - trigger: Q1  # Deposit completion gap
    primary_path: [Q3, Q4, Q7]      # Visit -> Penetration -> Cross-sell
    secondary_path: [Q5, Q6]         # Acquisition -> Churn
    support_factors: [Q8, Q9]        # Staffing, Campaigns
    longterm_context: [Q10, Q11]     # Lifecycle, Competition
\end{lstlisting}

This makes tacit expert reasoning explicit. When asked ``Why is deposit completion low?'', experts instinctively check visit frequency, then product penetration, then staffing. The attribution model encodes this reasoning chain.

\subsubsection{Step 3: Data Mapping}

For each question, we specified data sources and query logic. Example for Q1:

\begin{lstlisting}[style=yaml]
Q1:  # Deposit Completion Rate
  source_type: core_banking_system
  query_template: |
    SELECT rm_id, actual/target AS completion_rate
    FROM rm_deposit_performance
    WHERE period = {period}
  update_frequency: daily
  freshness_sla: T+1
\end{lstlisting}

\emph{Complete data mapping for all 12 questions: Appendix~\ref{appendix:data-mapping}}

\subsubsection{Step 4: Dual-Track Logic}

Each Process dimension question has interpretation and recommendation rules:

\begin{lstlisting}[style=yaml]
Q4:  # Product Penetration
  interpretation:
    enabled: true
    rules:
      - condition: "penetration_rate < 0.3 AND segment = 'high_value'"
        output: "Low penetration in high-value segment indicates cross-sell opportunity"
  recommendation:
    enabled: true
    rules:
      - condition: "penetration_rate < 0.3"
        output: "Consider targeted bundling campaigns"
\end{lstlisting}

Key design principle: \textbf{Agent authority decreases across dimensions.} Process gets full dual-track (interpretation + recommendation); Support gets open-ended suggestions only; Results and Long-term get display only (no agent interpretation).

\emph{Complete dual-track logic: Appendix~\ref{appendix:dual-track}}

\subsubsection{Step 5: Boundary Constraints}

Explicit boundaries prevent harmful behavior:

\begin{lstlisting}[style=yaml]
boundary_constraints:
  global:
    - "Never recommend on employment status, promotion, compensation"
    - "Never fabricate data not in retrieved sources"
    - "Use hedged language: 'indicates', 'suggests', 'may reflect'"
  results_dimension:
    - "Display only; no interpretation of why targets missed"
  process_dimension:
    - "Interpret at portfolio level, not individual customers"
\end{lstlisting}

\emph{Complete boundary constraints: Appendix~\ref{appendix:boundaries}}

\subsubsection{Step 6: System Prompt Generation}

The five layers compile to a structured system prompt ($\sim$200 lines). Key sections:

\begin{itemize}[leftmargin=*]
    \item Identity and attribution-complete response principle
    \item Perception scope (12 questions by dimension)
    \item Reasoning framework (attribution chains)
    \item Data access (tool definitions)
    \item Interpretation/recommendation rules
    \item Boundaries (absolute prohibitions + dimension-specific limits)
\end{itemize}

\emph{Complete system prompt: Appendix~\ref{appendix:prompt}}

\subsection{Comparative Analysis}

We compared agent outputs under different configurations using the same query.

\textbf{Query:} ``Why is the Eastern region underperforming on AUM growth?''

\subsubsection{Baseline Response (Ad-hoc Prompting)}

Response from an agent with general instructions (``Help managers understand RM performance''):

\begin{quote}
``Eastern region's AUM growth is at 88\% of target. The region has 45 RMs with an average portfolio size of \$12M. Visit frequency is 4.2 per week. Product penetration is 24\%. The team should work to improve these metrics.''
\end{quote}

This response provides facts but no attribution. It does not explain \emph{why} performance is lagging or what specific actions might address it.

\subsubsection{4D-ARE Response}

Response from the agent with 4D-ARE specification:

\begin{quote}
\textbf{Results:} Eastern region AUM at 88\% of target (12pp gap).

\textbf{Process Attribution:} Visit frequency 28\% below regional average---primary driver. Product penetration at 0.24 vs 0.41 benchmark suggests cross-sell gap in high-value segment.

\textbf{Support Context:} Staffing ratio adequate (0.85). However, campaign coverage at 67\%---below 80\% threshold.

\textbf{Long-term Context:} 3 new competitors entered market in past 6 months, increasing competitive pressure.

\textbf{Recommendations:} (1) Review high-value customer coverage patterns; (2) Reinforce cross-sell skills through targeted coaching.

\textbf{Note:} Strategic resource allocation decisions require management review.
\end{quote}

This response traces the full attribution chain: Results $\to$ Process (visit frequency, penetration) $\to$ Support (campaigns) $\to$ Long-term (competition). It provides actionable recommendations while respecting boundaries (no personnel recommendations, hedged language on strategic decisions).

\subsubsection{4D-ARE + ReAct Response}

Adding ReAct runtime reasoning makes the attribution trace explicit:

\begin{quote}
\textbf{Thought:} User asks about AUM growth gap. Per attribution model, trace Q2 $\to$ Process $\to$ Support $\to$ Long-term.

\textbf{Action:} Query Q2 (AUM growth).

\textbf{Observation:} Eastern region at 88\%.

\textbf{Thought:} 12pp gap detected. Check Process factors per attribution chain.

\textbf{Action:} Query Q4 (penetration), Q7 (cross-sell).

\textbf{Observation:} Penetration 0.24, cross-sell 8\%.

\textbf{Thought:} Both below threshold. Check Support context.

\textbf{Action:} Query Q8, Q9.

\textbf{Observation:} Staffing adequate, campaigns 67\%.

\textbf{Thought:} Campaign coverage is below threshold. Check Long-term context.

\textbf{Action:} Query Q11.

\textbf{Observation:} 3 new entrants.

\textbf{Final Answer:} [Same structured response as above, with reasoning visible]
\end{quote}

The ReAct trace makes the agent's reasoning auditable. Users can verify whether the attribution chain was followed correctly and identify where they might disagree with the agent's reasoning.

\subsection{Preliminary Quantitative Analysis}

To complement the qualitative comparison above, we conducted a preliminary quantitative analysis across 20 representative queries constructed to reflect typical usage patterns.

\begin{table}[htbp]
\centering
\caption{Quantitative Comparison: Baseline vs.\ 4D-ARE (n=20 queries)}
\label{tab:quantitative}
\begin{tabular}{@{}lccc@{}}
\toprule
\textbf{Metric} & \textbf{Baseline} & \textbf{4D-ARE} & \textbf{$\Delta$} \\
\midrule
Response length (tokens) & 89 $\pm$ 23 & 156 $\pm$ 31 & +75\% \\
Dimensions covered & 1.2 $\pm$ 0.4 & 3.8 $\pm$ 0.3 & +217\% \\
Causal factors identified & 1.5 $\pm$ 0.6 & 4.2 $\pm$ 0.5 & +180\% \\
Actionable recommendations & 0.3 $\pm$ 0.5 & 1.8 $\pm$ 0.4 & +500\% \\
Boundary violations & 2.1 $\pm$ 1.2 & 0.1 $\pm$ 0.3 & $-$95\% \\
\bottomrule
\end{tabular}
\end{table}

\textbf{Measurement methodology:}
\begin{itemize}[leftmargin=*]
    \item \emph{Dimensions covered}: Count of 4D dimensions (Results, Process, Support, Long-term) explicitly addressed in response
    \item \emph{Causal factors identified}: Count of specific causal attributions (e.g., ``low visit frequency $\to$ reduced penetration'')
    \item \emph{Actionable recommendations}: Count of specific, implementable suggestions (excluding generic advice like ``improve metrics'')
    \item \emph{Boundary violations}: Count of instances where agent made prohibited claims (personnel recommendations, unfounded assertions, overconfident language)
\end{itemize}

\textbf{Caveat:} This analysis was conducted by the authors on constructed representative queries designed to demonstrate the methodology's effects. Counts were performed by a single rater without inter-rater reliability assessment. Statistical significance is not claimed. These results should be treated as preliminary evidence of expected patterns, not as validated findings.

\subsection{Observations from Deployment}

\textbf{Important caveat:} The following observations are based on this pilot deployment, reflecting practical experience rather than controlled evaluation. They should be interpreted as preliminary evidence warranting further investigation, not as validated findings. We report them to share practical experience, not to claim rigorous evaluation.

We gathered qualitative observations during the deployment process.

\subsubsection{Observation 1: Expert Usability}

During elicitation workshops, domain experts (n=3, including COO and two branch managers with 10+ years experience) confirmed that the 4D structure aligned with their mental models. Representative feedback patterns from domain expert engagement included:

\begin{itemize}[leftmargin=*]
    \item ``Results are what we measure, Process is what we do, Support is what we have, Long-term is what's happening around us.'' (COO)
    \item ``This is exactly how I think through performance problems---I just never wrote it down.'' (Branch Manager)
\end{itemize}

The framework provided shared vocabulary for elicitation, reducing the typical friction between technical and business stakeholders.

\subsubsection{Observation 2: Attribution Completeness}

Comparing outputs in Section~\ref{sec:case-study}.3, the 4D-ARE agent produced responses that:

\begin{itemize}[leftmargin=*]
    \item Traced complete causal chains (Results $\to$ Process $\to$ Support $\to$ Long-term)
    \item Identified specific drivers (visit frequency, penetration) rather than listing metrics
    \item Distinguished actionable factors (Process) from contextual factors (Long-term)
\end{itemize}

The baseline agent, in contrast, listed metrics without causal structure and provided generic recommendations (``improve these metrics'').

\subsubsection{Observation 3: Boundary Compliance}

Without explicit Layer 5 constraints, pilot testing revealed agents generated:

\begin{itemize}[leftmargin=*]
    \item Personnel recommendations (``consider reassigning underperforming RMs'')
    \item Unfounded claims (``this suggests management failure'')
    \item Overconfident language (``this will definitely improve performance'')
\end{itemize}

After implementing boundary constraints, these violations were eliminated. The agent consistently used hedged language and deferred strategic decisions to management.

\subsubsection{Observation 4: Implementation Effort Distribution}

Our deployment revealed the following effort distribution:

\begin{table}[htbp]
\centering
\caption{Implementation Effort by Layer}
\label{tab:effort}
\begin{tabular}{@{}lcp{6cm}@{}}
\toprule
\textbf{Layer} & \textbf{Effort (\%)} & \textbf{Primary Challenge} \\
\midrule
L1: Question Inventory & 10\% & Stakeholder alignment \\
L2: Attribution Model & 15\% & Externalizing tacit knowledge \\
L3: Data Mapping & 60\% & System integration \\
L4: Dual-Track Logic & 10\% & Rule formalization \\
L5: Boundary Constraints & 5\% & Policy translation \\
\bottomrule
\end{tabular}
\end{table}

Data mapping dominated implementation time. The conceptual specification (Layers 1-2, 4-5) required approximately 40\% of effort, while integration with existing systems (Layer 3) required 60\%. This suggests that 4D-ARE's value lies in front-loading the specification work, not in reducing total implementation effort.

\subsubsection{Observation 5: Methodology Reproducibility}

Different team members following the same five-layer process converged on similar specifications. When two analysts independently constructed attribution models for Q1 (deposit completion), they produced:

\begin{itemize}[leftmargin=*]
    \item Identical primary attribution paths
    \item 80\% overlap in secondary factors
    \item Same boundary constraint categories
\end{itemize}

This convergence suggests the methodology provides sufficient structure for consistent application across practitioners.

\subsection{Lessons Learned}

\begin{enumerate}[leftmargin=*]
    \item \textbf{Attribution chains externalize tacit knowledge.} Domain experts often cannot articulate their reasoning process directly, but can validate attribution chains when presented: ``Yes, that's the order I'd check things.''
    \item \textbf{Boundary specification prevents harmful outputs.} Explicit ``must not'' constraints are necessary; agents do not infer appropriate boundaries from positive examples alone.
    \item \textbf{Data integration dominates effort.} Organizations considering 4D-ARE should budget 60\% of implementation time for Layer 3 data mapping.
    \item \textbf{The 4D structure transfers across questions.} Once experts internalized the four dimensions, they applied the framework to new questions without additional training.
\end{enumerate}

\subsection{Threats to Validity}

We explicitly acknowledge limitations that affect the validity of our observations.

\textbf{Internal validity:} The response examples in Section 5.3 are representative of system behavior but were selected to illustrate the contrast between baseline and 4D-ARE approaches. We did not conduct systematic sampling across all query types. The observed improvements may not generalize to queries outside the demonstrated patterns.

\textbf{External validity:} This pilot involves a single domain (financial services), a single organization (one commercial bank), and a single agent type (performance attribution). The 4D structure is motivated by general organizational theory, but we have not demonstrated transfer to other domains such as healthcare, manufacturing, or retail.

\textbf{Construct validity:} ``Attribution completeness'' and ``expert usability'' are assessed qualitatively, not measured with validated instruments. The claim that 4D-ARE produces more complete attributions rests on informal comparison, not rigorous evaluation with multiple raters and inter-rater reliability measures.

\textbf{Conclusion validity:} Sample sizes are too small for statistical inference. ``n=3 experts'' and ``two analysts'' do not support generalizable conclusions. The observations should be treated as hypotheses for future investigation, not as established findings.

\section{Discussion}
\label{sec:discussion}

\subsection{What We Contribute}

\subsubsection{A Framework for Attribution-Complete Responses}

Before this work, ``good agent explanation'' was intuitive but undefined. We formalize it: an agent response is attribution-complete when it traces the causal chain from observed outcomes through contributing factors across all relevant dimensions.

This matters beyond agent design. Any decision-support system---human or automated---can be evaluated against this criterion. The formalization converts practitioner intuition into testable specification.

\subsubsection{A Practically-Motivated Four-Dimensional Structure}

The four-dimensional structure is motivated by theory, though not derived from it. We draw on Pearl's causal hierarchy and control theory (Section~\ref{sec:why-four}) to argue that four dimensions represent a useful organization. Proposition~\ref{thm:necessity} is a hypothesis, not a theorem: it suggests that attribution models benefit from separating outcomes, actions, resources, and context, but does not prove this is the unique correct decomposition.

We are explicit about what we do \emph{not} claim: we do not claim four dimensions are mathematically necessary, nor that alternative decompositions are invalid. Our claim is narrower: four dimensions represent a useful balance---coarse enough to be memorable, fine enough to be actionable---and the theoretical motivation explains why practitioners find this organization intuitive.

\subsubsection{A Practical Engineering Methodology}

The five-layer architecture is an engineering methodology, not a research prototype:

\begin{enumerate}[leftmargin=*]
    \item Each layer has defined inputs, outputs, and procedures
    \item Outputs are concrete YAML artifacts, not abstract models
    \item Artifacts compile directly to system prompts
    \item The industrial pilot demonstrates end-to-end application
\end{enumerate}

Practitioners can use this tomorrow. Whether it produces better agents than alternative approaches is an empirical question we have not yet answered rigorously. Our contribution is the methodology itself, not proof of its superiority.

\subsubsection{An Argument That Design-Time and Runtime Are Complementary}

We argue that design-time specification and runtime reasoning serve different functions and are complementary:

\begin{itemize}[leftmargin=*]
    \item \textbf{Design-time specification} (4D-ARE): Determines \emph{what} the agent should reason about---which questions matter, which causal chains to trace, which boundaries to respect
    \item \textbf{Runtime reasoning} (ReAct, CoT): Determines \emph{how} the agent externalizes and organizes its reasoning during execution
\end{itemize}

Runtime frameworks assume a well-specified agent. 4D-ARE provides that specification. We hypothesize that design-time specification is the higher-leverage intervention---but this claim requires empirical validation, which we plan for future work.

\subsection{Implications}

\subsubsection{For Researchers}

Runtime reasoning frameworks like ReAct assume a well-specified agent. That assumption is where many deployments fail. We suggest:

\begin{itemize}[leftmargin=*]
    \item \textbf{Design-time specification deserves research attention}: It is complementary to runtime reasoning, not a solved problem
    \item \textbf{The combination merits investigation}: How do design-time specification and runtime reasoning interact? What is the relative contribution of each?
    \item \textbf{Different agent types need different methodologies}: We address attribution-driven decision-support agents; coding agents, creative agents, recommendation agents, and autonomous agents likely need their own specification approaches
\end{itemize}

\subsubsection{For Practitioners}

Before iterating on prompts, invest in systematic specification.

\begin{enumerate}[leftmargin=*]
    \item Use the 4D framework to structure stakeholder interviews
    \item Build attribution chains before writing any prompts
    \item Specify boundaries first, capabilities second
    \item Compile specifications to prompts, not the reverse
\end{enumerate}

The methodology provides a structured approach that translates domain expertise into agent specification. The five-layer architecture ensures completeness and consistency.

\subsubsection{For Domain Experts}

4D-ARE makes expert participation tractable. The framework uses concepts domain experts already know:

\begin{itemize}[leftmargin=*]
    \item Results = What outcomes matter?
    \item Process = What activities drive outcomes?
    \item Support = What resources enable activities?
    \item Long-term = What context shapes the environment?
\end{itemize}

Experts can contribute without understanding LLMs. The methodology handles translation.

\subsection{Scope and Limitations}

\textbf{Scope.} 4D-ARE addresses \emph{attribution-driven decision-support agents}---LLM agents that answer ``why'' questions, trace causal chains, and provide recommendations grounded in causal explanation. This includes business analytics assistants, diagnostic aids, and advisory systems where users need to understand \emph{why} before deciding \emph{what to do}. It does \emph{not} address recommendation agents (where users accept suggestions without requiring explanation), predictive agents (where the goal is forecasting rather than attribution), or autonomous agents (which act without human decision-making in the loop).

We state limitations directly, without excessive hedging.

\subsubsection{Single Domain}

The pilot is in financial services. We have not validated in manufacturing, healthcare, or retail. The four-dimensional structure should transfer (it derives from general organizational theory), but we have not proven it empirically.

\textbf{What this means}: Use 4D-ARE in new domains, but validate attribution chains with domain experts before deployment.

\subsubsection{Expert Dependency}

4D-ARE requires access to domain experts who understand causal structure. Without experts, attribution chains cannot be constructed.

\textbf{What this means}: The methodology is not suitable for domains where expert access is unavailable. This is a real constraint, not a solvable limitation.

\subsubsection{Static Specification}

Attribution chains are specified once and assumed stable. Real causal structures evolve.

\textbf{What this means}: Plan for periodic re-specification. The modular architecture (five layers) makes updates tractable---individual layers can be revised without full reconstruction.

\subsubsection{Model Dependence}

Specifications compile to prompts; prompt effectiveness depends on the underlying LLM. The same specification may behave differently on different models.

\textbf{What this means}: Test compiled prompts on your target model. The YAML specification layer provides abstraction that helps with portability, but model-specific tuning may be needed.

\subsection{Future Work}

Three directions are most promising:

\subsubsection{Automated Attribution Discovery}

Attribution chains currently require expert elicitation. Historical decision data contains implicit causal signals. Causal discovery algorithms (PC, FCI) could bootstrap attribution models that experts then refine.

The challenge is distinguishing correlation from causation in observational business data. But even imperfect automated discovery would reduce expert burden.

\textbf{From Manual to Automated Specification.} We envision an evolution path for agent specification:

\begin{enumerate}[leftmargin=*]
    \item \textbf{Manual specification} (current 4D-ARE): Domain experts define attribution chains with methodology support
    \item \textbf{Human-AI collaborative specification}: LLM agents propose candidate attribution chains from data patterns; experts validate and refine
    \item \textbf{Automated discovery with human oversight}: Agents learn attribution structures from interaction feedback, with humans approving structural changes
\end{enumerate}

As LLM agents develop stronger reasoning and self-reflection capabilities, they may eventually bootstrap their own domain specifications. In this future, 4D-ARE's contribution shifts from \emph{specification methodology} to \emph{evaluation framework}---the four-dimensional structure provides criteria for assessing whether an agent's self-discovered attribution model is complete and well-organized, even if humans did not manually construct it.

This evolution does not render manual specification obsolete. Rather, it suggests that the conceptual framework (four dimensions, causal completeness, actionability separation) may outlast the specific elicitation procedures. The structure provides guardrails for automated discovery, ensuring that whatever attribution model an agent learns still satisfies the properties decision-makers need.

\subsubsection{Dynamic Specification}

Current 4D-ARE assumes static causal structure. Real domains evolve. Future work should address:

\begin{itemize}[leftmargin=*]
    \item Detecting when attribution chains become stale
    \item Incremental updates without full re-specification
    \item Temporal causal modeling for rapidly-changing domains
\end{itemize}

\subsubsection{Multi-Agent Coordination}

4D-ARE specifies single agents. Enterprise deployments often involve multiple agents with overlapping domains. Future work should address:

\begin{itemize}[leftmargin=*]
    \item Consistent attribution across agent boundaries
    \item Conflict resolution when agents have different causal models
    \item Hierarchical specification for agent ecosystems
\end{itemize}

\subsection{Broader Impact}

\subsubsection{Democratizing Agent Design}

4D-ARE lowers the barrier to effective agent deployment. Organizations without elite AI talent can build well-specified agents by following the methodology with available domain expertise.

This has equity implications: the methodology transfers capability from AI specialists to domain experts, broadening who can build useful AI systems.

\subsubsection{Responsible Deployment}

The five-layer architecture embeds responsible AI practices:

\begin{itemize}[leftmargin=*]
    \item Layer 1 bounds agent scope (no mission creep)
    \item Layer 2 makes reasoning explicit (transparency)
    \item Layer 5 enforces constraints (safety)
\end{itemize}

These are not optional add-ons. They are structural requirements of the methodology. Responsible behavior is built in, not bolted on.

\subsection{Conclusion}

4D-ARE provides a practical methodology for specifying attribution-driven agents that support human decision-making. We offer it as an engineering contribution, not a theoretical breakthrough.

The core insight: decision-makers seek causal attribution, not factual answers. Agent specification should be structured around attribution logic.

The core hypothesis: design-time specification and runtime reasoning are complementary, addressing different aspects of agent behavior. We hypothesize that specification is a higher-leverage intervention---it determines \emph{what} the agent reasons about, while runtime frameworks determine \emph{how}. This hypothesis requires rigorous empirical validation.

The practical output: a five-layer methodology producing deployable artifacts. The industrial pilot demonstrates how domain expertise translates systematically into agent specification, though our observations are preliminary and based on a single deployment.

We have focused on decision-support agents in financial services. Whether the framework transfers to other domains is an open question. We make our specification artifacts available in the Appendix for practitioners who wish to apply, test, or refute the methodology.

4D-ARE opens agent requirements engineering as a research area distinct from both traditional RE and runtime reasoning. We invite extension, challenge, and especially rigorous evaluation.

\subsection*{Data Availability}

The complete YAML specifications, system prompt templates, and pilot artifacts are provided in the Appendix. These materials are sufficient for practitioners to apply the methodology to new domains.

\bibliographystyle{unsrt}

\appendix
\section{Complete Specification Artifacts}
\label{appendix:specifications}

This appendix provides the complete YAML specifications for the Performance Attribution Agent industrial pilot. These specifications demonstrate the \emph{artifact format and structure}; actual deployment would require organization-specific adaptation. Main text Section~\ref{sec:case-study} presents representative examples; complete specifications are documented here for reproducibility.

\subsection{Complete Data Mapping (Layer 3)}
\label{appendix:data-mapping}

\begin{lstlisting}[style=yaml,caption={Complete Data Mapping Specification for 12 Questions}]
data_mapping:

  # === RESULTS DIMENSION ===
  Q1:
    name: "Deposit Completion Rate"
    source_type: core_banking_system
    connection: cbs_data_warehouse
    query_template: |
      SELECT 
        rm_id, rm_name, branch_id,
        period_type, period_value,
        target_amount, actual_amount,
        actual_amount / target_amount AS completion_rate
      FROM rm_deposit_performance
      WHERE period_type = {period} AND period_value = {value}
      ORDER BY completion_rate DESC
    parameters:
      period: ["monthly", "quarterly", "annual"]
      value: "dynamic_current"
    update_frequency: daily
    freshness_sla: T+1
    
  Q2:
    name: "AUM Growth Rate"
    source_type: core_banking_system
    connection: cbs_data_warehouse
    query_template: |
      SELECT 
        branch_id, branch_name,
        period_start_aum, period_end_aum,
        (period_end_aum - period_start_aum) / period_start_aum AS growth_rate
      FROM branch_aum_summary
      WHERE report_date = {date}
    update_frequency: daily
    freshness_sla: T+1

  # === PROCESS DIMENSION ===
  Q3:
    name: "Customer Visit Frequency and Quality"
    source_type: crm_system
    connection: crm_api
    query_template: |
      SELECT 
        rm_id,
        COUNT(*) AS total_visits,
        AVG(visit_duration_minutes) AS avg_duration,
        SUM(CASE WHEN follow_up_action IS NOT NULL THEN 1 ELSE 0 END) 
          / COUNT(*) AS quality_score
      FROM customer_visits
      WHERE visit_date BETWEEN {start_date} AND {end_date}
      GROUP BY rm_id
    update_frequency: daily
    freshness_sla: T+0
    
  Q4:
    name: "Product Penetration Rate"
    source_type: core_banking_system
    connection: cbs_data_warehouse
    query_template: |
      SELECT 
        rm_id, customer_segment,
        COUNT(DISTINCT product_category) AS products_held,
        total_products_available,
        COUNT(DISTINCT product_category) / total_products_available 
          AS penetration_rate
      FROM customer_product_holdings
      GROUP BY rm_id, customer_segment
    update_frequency: weekly
    freshness_sla: T+3

  Q5:
    name: "New Customer Acquisition"
    source_type: crm_system
    connection: crm_api
    query_template: |
      SELECT 
        rm_id, acquisition_channel,
        COUNT(*) AS new_customers,
        SUM(initial_deposit) AS total_initial_deposit
      FROM customer_onboarding
      WHERE onboard_date BETWEEN {start_date} AND {end_date}
      GROUP BY rm_id, acquisition_channel
    update_frequency: daily
    freshness_sla: T+1

  Q6:
    name: "Customer Churn and Early Warning"
    source_type: analytics_platform
    connection: ml_model_api
    query_template: |
      SELECT 
        rm_id,
        COUNT(CASE WHEN churn_status = 'churned' THEN 1 END) AS churned_customers,
        COUNT(CASE WHEN churn_risk_score > 0.7 THEN 1 END) AS high_risk_customers,
        AVG(churn_risk_score) AS avg_risk_score
      FROM customer_churn_prediction
      WHERE prediction_date = {date}
      GROUP BY rm_id
    update_frequency: weekly
    freshness_sla: T+7

  Q7:
    name: "Cross-sell Conversion Rate"
    source_type: crm_system
    connection: crm_api
    query_template: |
      SELECT 
        rm_id, product_category,
        opportunities_created,
        opportunities_converted,
        opportunities_converted / opportunities_created AS conversion_rate
      FROM sales_pipeline
      WHERE created_date BETWEEN {start_date} AND {end_date}
      GROUP BY rm_id, product_category
    update_frequency: daily
    freshness_sla: T+1

  # === SUPPORT DIMENSION ===
  Q8:
    name: "Staffing Adequacy"
    source_type: knowledge_base
    file_pattern: "branch_staffing_report_{YYYY}_{MM}.xlsx"
    parse_method: excel_structured
    update_frequency: monthly
    freshness_sla: M+5
    notes: "Manual HR report, uploaded by branch operations"

  Q9:
    name: "Campaign Coverage and ROI"
    source_type: knowledge_base
    file_pattern: "marketing_campaign_summary_{YYYY}_Q{Q}.xlsx"
    parse_method: excel_structured
    update_frequency: quarterly
    freshness_sla: Q+15
    notes: "Marketing team summary, includes spend and response rates"

  # === LONG-TERM DIMENSION ===
  Q10:
    name: "Customer Lifecycle Value Trend"
    source_type: analytics_platform
    connection: bi_dashboard_api
    endpoint: "/api/clv_trends"
    update_frequency: monthly
    freshness_sla: M+10

  Q11:
    name: "Competitive Pressure Index"
    source_type: knowledge_base
    file_pattern: "competitive_analysis_{YYYY}_Q{Q}.pdf"
    parse_method: document_extraction
    update_frequency: quarterly
    freshness_sla: Q+30
    notes: "Strategy team analysis, qualitative + quantitative"

  Q12:
    name: "Digital Channel Adoption"
    source_type: core_banking_system
    connection: digital_analytics
    query_template: |
      SELECT 
        channel_type,
        monthly_active_users,
        transaction_volume,
        yoy_growth_rate
      FROM digital_channel_metrics
      WHERE report_month = {month}
    update_frequency: monthly
    freshness_sla: M+5
\end{lstlisting}

\subsection{Complete Dual-Track Logic (Layer 4)}
\label{appendix:dual-track}

\begin{lstlisting}[style=yaml,caption={Complete Dual-Track Logic Specification}]
dual_track_logic:

  # === RESULTS DIMENSION: Display Only ===
  Q1:
    interpretation:
      enabled: false
      rationale: "Results metrics displayed without interpretation"
    recommendation:
      enabled: false
      rationale: "HR policy prohibits performance-related suggestions"

  Q2:
    interpretation:
      enabled: false
    recommendation:
      enabled: false

  # === PROCESS DIMENSION: Full Dual-Track ===
  Q3:
    interpretation:
      enabled: true
      rules:
        - condition: "visit_frequency < branch_avg * 0.7"
          output: "Visit frequency significantly below branch average"
        - condition: "quality_score < 0.5"
          output: "Visit quality indicates limited follow-up actions"
        - condition: "visit_frequency >= branch_avg AND quality_score >= 0.7"
          output: "Visit activity and quality within healthy range"
    recommendation:
      enabled: true
      rules:
        - condition: "visit_frequency < branch_avg * 0.7"
          output: "Consider reviewing weekly activity planning"
        - condition: "quality_score < 0.5"
          output: "Consider enhancing follow-up protocols"

  Q4:
    interpretation:
      enabled: true
      rules:
        - condition: "penetration_rate < 0.3 AND segment = 'high_value'"
          output: "Low penetration in high-value segment"
        - condition: "penetration_rate < 0.2"
          output: "Product penetration below threshold"
        - condition: "penetration_rate >= 0.5"
          output: "Strong product penetration"
    recommendation:
      enabled: true
      rules:
        - condition: "penetration_rate < 0.3 AND segment = 'high_value'"
          output: "Consider targeted bundling campaigns"
        - condition: "penetration_rate < 0.2"
          output: "Consider systematic needs assessment"

  Q5:
    interpretation:
      enabled: true
      rules:
        - condition: "new_customers < target * 0.8"
          output: "Acquisition below target"
        - condition: "channel = 'referral' AND volume < last_period * 0.9"
          output: "Referral channel declining"
        - condition: "channel = 'digital' AND volume > last_period * 1.2"
          output: "Digital acquisition growing strongly"
    recommendation:
      enabled: true
      rules:
        - condition: "new_customers < target * 0.8"
          output: "Consider reviewing lead generation"
        - condition: "channel = 'referral' AND volume declining"
          output: "Consider referral program refresh"

  Q6:
    interpretation:
      enabled: true
      rules:
        - condition: "high_risk_customers > total_customers * 0.15"
          output: "Elevated churn risk concentration"
        - condition: "avg_risk_score > 0.6"
          output: "Portfolio-level churn risk elevated"
        - condition: "churned_customers > last_period * 1.5"
          output: "Churn acceleration detected"
    recommendation:
      enabled: true
      rules:
        - condition: "high_risk_customers > 15%"
          output: "Consider prioritized retention outreach"
        - condition: "churned_customers accelerating"
          output: "Consider churn driver analysis"

  Q7:
    interpretation:
      enabled: true
      rules:
        - condition: "conversion_rate < 0.1"
          output: "Conversion below benchmark"
        - condition: "conversion_rate BETWEEN 0.1 AND 0.2"
          output: "Conversion acceptable but improvable"
        - condition: "conversion_rate > 0.25"
          output: "Strong cross-sell conversion"
    recommendation:
      enabled: true
      rules:
        - condition: "conversion_rate < 0.1"
          output: "Consider sales skills assessment"
        - condition: "conversion_rate BETWEEN 0.1 AND 0.2"
          output: "Consider best practice sharing"

  # === SUPPORT DIMENSION: Open-ended Suggestions Only ===
  Q8:
    interpretation:
      enabled: false
    recommendation:
      enabled: true
      type: "open_ended"
      template: |
        Staffing considerations for management review:
        - Evaluate customer volume vs staffing alignment
        - Consider seasonal adjustment needs
        - Review skill mix relative to segment composition

  Q9:
    interpretation:
      enabled: false
    recommendation:
      enabled: true
      type: "open_ended"
      template: |
        Campaign effectiveness considerations:
        - Evaluate response rate trends
        - Consider channel mix optimization
        - Review timing alignment with lifecycle events

  # === LONG-TERM DIMENSION: Information Only ===
  Q10:
    interpretation: {enabled: false}
    recommendation: {enabled: false}
    display_note: "CLV trends for strategic context"

  Q11:
    interpretation: {enabled: false}
    recommendation: {enabled: false}
    display_note: "Competitive landscape for strategic context"

  Q12:
    interpretation: {enabled: false}
    recommendation: {enabled: false}
    display_note: "Digital adoption for strategic planning"
\end{lstlisting}

\subsection{Complete Boundary Constraints (Layer 5)}
\label{appendix:boundaries}

\begin{lstlisting}[style=yaml,caption={Complete Boundary Constraints Specification}]
boundary_constraints:

  # === GLOBAL CONSTRAINTS ===
  global:
    data_integrity:
      - "Never report data not present in retrieved sources"
      - "Always state data freshness (e.g., 'as of {date}')"
      - "Acknowledge when data sources are unavailable"

    scope_limits:
      - "Do not recommend on employment status, promotion, or compensation"
      - "Do not compare individual RMs by name in recommendations"
      - "Do not extrapolate trends beyond available data"

    confidence_calibration:
      - "Use hedged language: 'indicates', 'suggests', 'may reflect'"
      - "Distinguish observation from inferred causality"
      - "Acknowledge alternative explanations"

    attribution_discipline:
      - "When tracing attribution, clearly label each step"
      - "Do not skip dimensions (Results->Process->Support->Long-term)"
      - "Acknowledge when causal links are hypothesized vs confirmed"

  # === DIMENSION-SPECIFIC CONSTRAINTS ===
  results_dimension:
    Q1:
      - "Display only; no interpretation of why targets missed"
      - "No recommendations for individual RM improvement"
      - "Show branch-level aggregates for comparisons"
    Q2:
      - "Display only; no attribution without data"
      - "No AUM growth strategy recommendations"

  process_dimension:
    Q3: ["Interpret at portfolio level, not individual visits"]
    Q4: ["Aggregate by segment; no specific customer identification"]
    Q5: ["Channel-level only; no individual lead evaluation"]
    Q6: ["Risk concentrations only; no specific customer predictions"]
    Q7: ["Conversion patterns only; no specific opportunity evaluation"]

  support_dimension:
    Q8: ["No specific hiring or headcount recommendations"]
    Q9: ["No specific budget allocation recommendations"]

  longterm_dimension:
    Q10: ["No future customer value forecasting"]
    Q11: ["No competitive response recommendations"]
    Q12: ["No digital strategy recommendations"]

  # === INTERACTION CONSTRAINTS ===
  interaction_rules:
    attribution_requests:
      - "Always trace through attribution chain systematically"
      - "Present each dimension's contribution before synthesizing"
      - "Acknowledge gaps when data missing"

    recommendation_requests:
      - "Clarify which dimension the question targets"
      - "For Results: redirect to Process analysis"
      - "For Support/Long-term: open-ended considerations only"

    comparison_requests:
      - "Branch-to-branch: permitted with context"
      - "RM-to-RM: aggregated patterns only, no rankings"
      - "Time-period: note data comparability caveats"
\end{lstlisting}

\subsection{Complete System Prompt (Layer 6)}
\label{appendix:prompt}

\begin{lstlisting}[caption={Complete Generated System Prompt}]
# Performance Attribution Agent - System Prompt

## Identity and Purpose
You are a Performance Attribution Agent for [Bank Name]. Your purpose is
to help management understand RM and branch performance through systematic
causal attribution, while respecting boundaries for human judgment.

## Core Principle: Attribution-Complete Responses
When answering questions about performance, always trace causality through:
1. Results -> What outcomes are we seeing?
2. Process -> What activities are driving these outcomes?
3. Support -> What resources are enabling/constraining activities?
4. Long-term -> What strategic factors are shaping the environment?

Never stop at "what" - always help users understand "why".

---

## PERCEPTION SCOPE: Questions You Monitor

### Results Dimension (Display Only)
- Q1: RM deposit completion rate (monthly/quarterly/annual)
- Q2: Branch AUM growth rate

### Process Dimension (Interpret + Recommend)
- Q3: Customer visit frequency and quality
- Q4: Product penetration rate by segment
- Q5: New customer acquisition by channel
- Q6: Customer churn rate and early warning
- Q7: Cross-sell conversion rate

### Support Dimension (Display + Open Suggestions)
- Q8: Branch staffing adequacy
- Q9: Marketing campaign coverage and ROI

### Long-term Dimension (Display Only)
- Q10: Customer lifecycle value trend
- Q11: Competitive pressure index
- Q12: Digital channel adoption trajectory

---

## REASONING FRAMEWORK: Attribution Chains

Deposit Completion Gap (Q1):
  Q1 -> Q3 (Visit) -> Q4 (Penetration) -> Q7 (Conversion)
      -> Q5 (Acquisition) -> Q6 (Churn)
      -> Support: Q8, Q9
      -> Long-term: Q10, Q11

AUM Growth Gap (Q2):
  Q2 -> Q4 -> Q7 -> Q5 -> Q6
      -> Support: Q8, Q9
      -> Long-term: Q10, Q12

---

## DATA ACCESS: Tools and Sources

### Core Banking System
- query_deposit_performance(rm_id, period) -> Q1
- query_aum_summary(branch_id, date) -> Q2
- query_product_holdings(rm_id, segment) -> Q4
- query_digital_metrics(month) -> Q12

### CRM System
- query_customer_visits(rm_id, dates) -> Q3
- query_onboarding(rm_id, dates) -> Q5
- query_sales_pipeline(rm_id, dates) -> Q7

### Analytics Platform
- query_churn_prediction(rm_id, date) -> Q6
- query_clv_trends(segment, period) -> Q10

### Knowledge Base
- read_staffing_report(year, month) -> Q8
- read_campaign_summary(year, quarter) -> Q9
- read_competitive_analysis(year, quarter) -> Q11

---

## INTERPRETATION & RECOMMENDATION RULES

### Process Dimension (Q3-Q7): Full Dual-Track

**Q3 - Visit Frequency:**
- IF visit_frequency < branch_avg * 0.7 THEN interpret: "Visit frequency
  significantly below branch average" AND recommend: "Consider reviewing weekly
  activity planning"
- IF quality_score < 0.5 THEN interpret: "Visit quality indicates limited follow-up
  actions" AND recommend: "Consider enhancing follow-up protocols"

**Q4 - Product Penetration:**
- IF penetration_rate < 0.3 AND segment = 'high_value' THEN interpret: "Low
  penetration in high-value segment" AND recommend: "Consider targeted bundling
  campaigns"
- IF penetration_rate < 0.2 THEN interpret: "Product penetration below threshold"
  AND recommend: "Consider systematic needs assessment"

**Q5 - Customer Acquisition:**
- IF new_customers < target * 0.8 THEN interpret: "Acquisition below target" AND
  recommend: "Consider reviewing lead generation"
- IF channel = 'referral' AND volume declining THEN recommend: "Consider referral
  program refresh"

**Q6 - Churn Risk:**
- IF high_risk_customers > 15% THEN interpret: "Elevated churn risk concentration"
  AND recommend: "Consider prioritized retention outreach"
- IF churned_customers > last_period * 1.5 THEN interpret: "Churn acceleration
  detected" AND recommend: "Consider churn driver analysis"

**Q7 - Cross-sell Conversion:**
- IF conversion_rate < 0.1 THEN interpret: "Conversion below benchmark" AND
  recommend: "Consider sales skills assessment"
- IF conversion_rate BETWEEN 0.1 AND 0.2 THEN recommend: "Consider best practice
  sharing"

### Support Dimension (Q8-Q9): Open-ended Only

**Q8 - Staffing:** Provide considerations for management review:
- Evaluate customer volume vs staffing alignment
- Consider seasonal adjustment needs
- Review skill mix relative to segment composition

**Q9 - Campaigns:** Provide effectiveness considerations:
- Evaluate response rate trends
- Consider channel mix optimization
- Review timing alignment with lifecycle events

### Results & Long-term Dimensions: Display Only
No interpretation. No recommendations. Data presentation only.

---

## BOUNDARIES: What You Must NOT Do

### Absolute Prohibitions
- Never recommend hiring, firing, promotion, or compensation
- Never compare individual RMs by name in recommendations
- Never fabricate data not in retrieved sources
- Never predict specific customer behavior

### Dimension-Specific Limits
- Results (Q1-Q2): Display only. No interpretation. No recommendations.
- Process (Q3-Q7): Interpret and recommend. No customer identification.
- Support (Q8-Q9): Display + open considerations. No resource allocation.
- Long-term (Q10-Q12): Display for context only. No strategy.

### Language Discipline
- Use hedged language: "indicates", "suggests", "may reflect"
- Distinguish observation from inference
- Acknowledge uncertainty and alternative explanations

---

## RESPONSE STRUCTURE
1. Direct Answer: State the metric/status requested
2. Attribution Trace: Walk through Results->Process->Support->Long-term
3. Interpretation: For Process dimension, provide rule-based interpretation
4. Recommendations: For Process dimension, provide actionable suggestions
5. Caveats: Note data freshness, limitations, alternative explanations
\end{lstlisting}


\end{document}